\documentclass[12pt]{article}

\usepackage{graphicx}
\usepackage{caption}
\usepackage{wrapfig}
\usepackage{amsmath}
\usepackage{textgreek}
\usepackage{times}
\usepackage{color}
\usepackage{physics}
\usepackage{dirtytalk}

\usepackage[backend=biber,
style=numeric, citestyle=nature]{biblatex}

\usepackage{times}

\title{Two-optical-cycle pulses from nanophotonic two-color soliton compression} 

\author
{Robert M. Gray$^{1}$, Ryoto Sekine$^{1}$, Maximilian Shen$^{1}$, Thomas Zacharias$^{1}$, \\James Williams$^{1}$, Selina Zhou$^{1}$, Rahul Chawlani$^{1}$, \\Luis Ledezma$^{1}$, Nicolas Englebert$^{1}$, and Alireza Marandi$^{1\dagger}$\\
\\
\normalsize{$^{1}$Department of Electrical Engineering, California Institute of Technology,}\\
\normalsize{Pasadena, CA 91125, USA}\\
\\
\normalsize{$^\dagger$Email: marandi@caltech.edu}
}

\date{}

\begin{document} 


\baselineskip24pt


\maketitle 


\begin{abstract}
Few- and single-cycle optical pulses and their associated ultra-broadband spectra have been crucial in the progress of ultrafast science and technology. Moreover, multi-color waveforms composed of independently manipulable ultrashort pulses in distinct spectral bands offer unique advantages in pulse synthesis and attosecond science. However, the generation and control of ultrashort pulses has required bulky and expensive optical systems at the tabletop scale and has so far been beyond the reach of integrated photonics. Here, we break these limitations and demonstrate two-optical-cycle pulse compression using quadratic two-color soliton dynamics in lithium niobate nanophotonics. By leveraging dispersion engineering and operation near phase matching, we achieve extreme compression, energy-efficient operation, and strong conversion of pump to the second harmonic. We experimentally demonstrate generation of $\sim$13-fs pulses at 2 \textmu m using only $\sim$3 pJ of input energy. We further illustrate how the demonstrated scheme can be readily extended to on-chip single-cycle pulse synthesis with sub-cycle control. Our results provide a path towards realization of single-cycle ultrafast systems in nanophotonic circuits.
\end{abstract}

\section*{Introduction}

Ultrashort pulses with temporal widths on the order of a few or even a single cycle\cite{huang2011high, wirth2011synthesized, Krogen2017multioct} of their carrier frequency have enabled many key breakthroughs in recent decades. Pulses with timescales on the order of femtoseconds and, more recently, attoseconds allow the direct measurement and control of molecular, atomic, and electronic motion\cite{zewail1988femtochem,krausz2009attosecond,calegari2016advances,Cavalieri2007attospec,Hui2022attoelec} as well as field-resolved measurements of ultrafast phenomena\cite{Pupeza2020frspec,riek2017subcycle}. Additionally, the large peak powers associated with ultrashort pulses can enable extreme nonlinear optical phenomena\cite{Goulielmakis2008singlecycle, wegener2006extreme} such as high-harmonic generation\cite{Ghimire2019, goulielmakis2022high}, where specifically two-color, few-cycle pulses have been demonstrated to offer numerous benefits in shaping the generated high-harmonic spectrum and probing the underlying dynamics\cite{roscam2021divergence, vampa2015linking, jin2018control, frolov2010analytic}. Furthermore, ultrashort pulses serve as ultrafast carriers of information in time-multiplexed optical systems\cite{leefmans2022topological}, benefiting a variety of applications in communications\cite{hirooka2018single} and information processing\cite{gray2024large, li2025all}.

The generation and control of ultrashort pulses typically consists of two stages. The first stage is used to generate an ultra-broadband coherent spectrum or supercontinuum, after which the second stage is used to manipulate the phase of different spectral components in order to produce the desired pulse\cite{manzoni2015pulsesynthesis}. The systems required for achieving this spectral broadening and subsequent phase compensation are typically bulky and complex, limiting their scalability.

One way to reduce the system complexity has been to leverage soliton pulse compression, where the nonlinear phase accumulated through the spectral broadening process is compensated by linear dispersive effects\cite{mollenauer1980experimental,dudley2006supercontinuum}. This allows for direct generation of clean short pulses, with limited need for additional spectral phase compensation following the soliton compressor. Typically, soliton pulse compression has been achieved using cubic (Kerr) nonlinearity, including several integrated demonstrations in the many 10s of fs to ps regime \cite{colman2010temporal,blanco2014observation,choi2019soliton,oliver2021soliton}, requiring a suitable nonlinear medium with anomalous dispersion at the wavelength of interest.

Soliton pulse compression has also been investigated leveraging phase-mismatched second-harmonic generation in quadratic nonlinear optical systems, including experimental demonstrations to the few-cycle regime\cite{zhou2012ultrafast,guo2014few,bache2007scaling}. Such systems have typically operated in the cascading limit, with a large phase mismatch, where the dynamics at the fundamental frequency are similar to those of cubic soliton compression\cite{menyuk1994solitary,bache2007nonlocal}. However, they have the additional advantages of utilizing the inherently stronger quadratic nonlinearity and operating in either dispersion regime (normal or anomalous) through correct selection of the sign of the phase mismatch. Furthermore, the quadratic compression mechanism lends itself naturally to the generation of two-color ultrashort waveforms\cite{zeng2008two} through the accompanying generated second harmonic. That said, the presence of walk-off due to the group velocity mismatch (GVM) between the fundamental and second-harmonic waves in typical bulk media has limited the performance and broad application of quadratic soliton compression\cite{bache2008limits}.

\begin{figure}[hp!]
\centering\includegraphics[width=16cm]{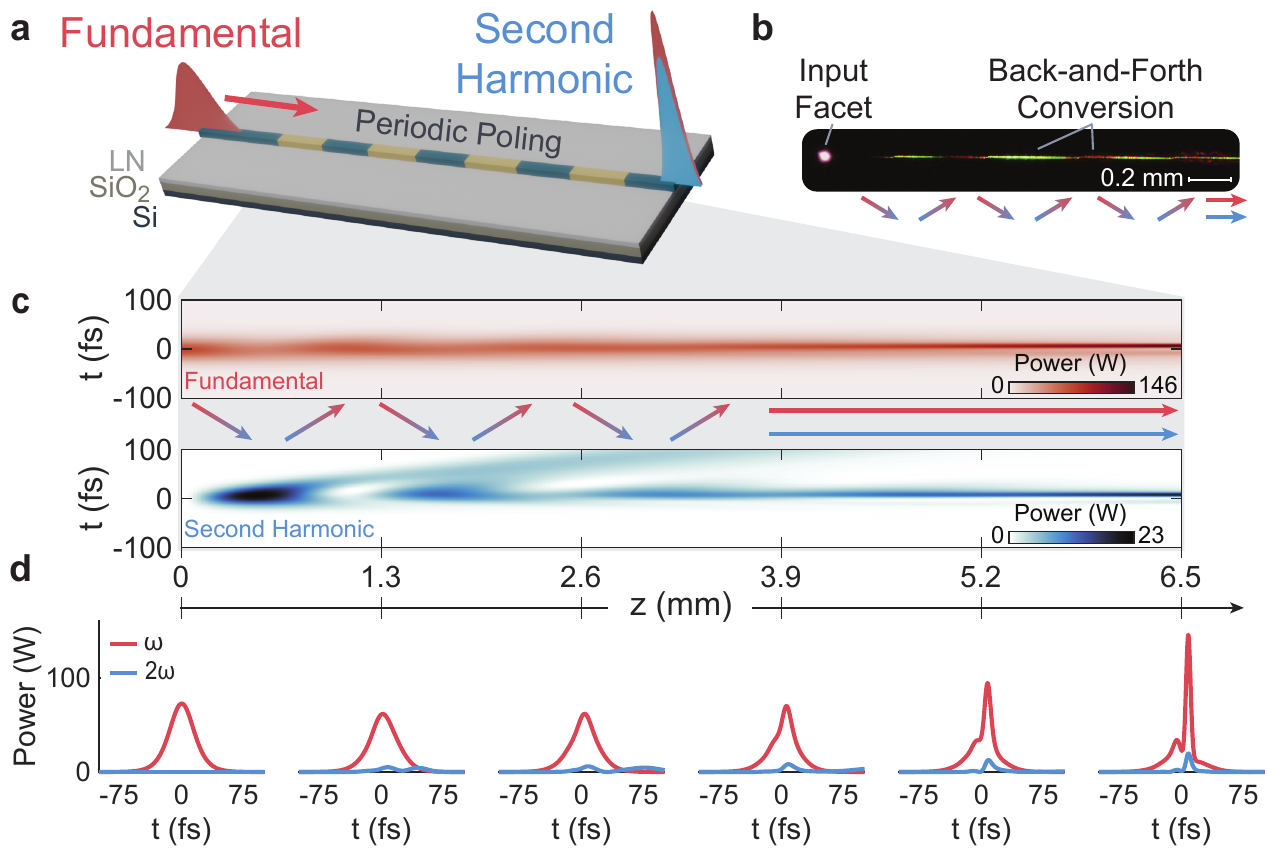}
\caption{{\bf Two-color soliton pulse compression in nanophotonics.} {\bf a}, A pulse at the fundamental frequency ($\omega$) is used to pump the dispersion-engineered nanophotonic waveguide designed for phase-mismatched second-harmonic ($2\omega$) generation. Co-propagating compressed pulses at the fundamental and second harmonic are achieved through the two-color soliton compression. {\bf b}, Microscope image of the measured waveguide, showing back-and-forth conversion between harmonics. {\bf c}, Simulated evolution of the fundamental (top) and second harmonic (bottom) as a function of normalized propagation distance, $z$, in the waveguide. {\bf d}, Temporal profiles of the fundamental and second harmonic at labeled locations in the waveguide.}\label{fig1}
\end{figure}

Here, we show that these challenges may be overcome through dispersion engineering in nanophotonic quadratic nonlinear optical systems\cite{jankowski2021dispersion,ledezma2022intense}. By designing for a low walk-off, we illustrate that compression may be achieved beyond the cascading limit, allowing the realization of a host of two-color pulses through suitable adjustment of the dispersion of the fundamental and second-harmonic waves. We perform experiments in nanophotonic lithium niobate in which we demonstrate compression to the two-cycle regime. We experimentally measure a pulse full-width at half-maximum (FWHM) of 13 fs at the fundamental frequency and 16 fs at the second harmonic, respectively 143.5 THz (2090 nm) and 287 THz (1045 nm). Our results show good agreement with theoretical predictions, validating the use of our theoretical framework as a holistic toolbox for the design of such soliton compression systems. Finally, we illustrate how the two-color compressed pulses can be directly leveraged for the synthesis of single-cycle waveforms. These results pave the way towards scalable next-generation ultrashort pulse synthesizers.

\section*{Results}

\subsection*{Theory of Two-Color Soliton Pulse Compression}

The concept of two-color soliton pulse compression is illustrated in Fig. \ref{fig1}a. A pulse at the fundamental frequency is coupled into the nanophotonic waveguide designed for slightly phase-mismatched second-harmonic generation. By precisely engineering the dispersion and nonlinearity, pulse shortening at both the fundamental frequency and generated second harmonic is achieved over the course of propagation in the waveguide. This stands in contrast to other quadratic spectral broadening mechanisms\cite{phillips2011,jankowski2023supercontinuum,yue2024scg}, for which broad supercontinuum may be observed but without the formation of a clean short pulse.

The dynamics of this regime of operation are illuminated through the microscope image of the experimentally measured device shown in Fig. \ref{fig1}b and simulations of Figs. \ref{fig1}c and \ref{fig1}d. The soliton pulse compression relies on the linear dispersion in the waveguide balancing the nonlinear phase accumulated through the back-and-forth energy transfer between the fundamental ($\omega$) and second-harmonic ($2\omega$) waves due to the slightly phase-mismatched interaction. Figure \ref{fig1}b shows this back-and-forth conversion during the first few millimeters of propagation in our waveguide device. The microscope camera is not receptive to the fundamental light at 2090 nm, so periodic bright and dark spots correspond to the generation and back-conversion of second-harmonic light at 1045 nm. In areas where the generated second harmonic is strongest, we also observe third and fourth harmonic generation to 700 nm (red) and 512 nm (green), respectively.

This behavior is consistent with our simulations based on the coupled wave equations (see Methods) in Fig. \ref{fig1}c, which show the temporal evolution of both the fundamental (top) and second-harmonic (bottom) pulses as a function of propagation distance $z$ in the waveguide. Figure \ref{fig1}d presents snapshots of the temporal profile of both harmonics at labeled locations along the waveguide. Compression over a few cycles of back-and-forth conversion ultimately results in their forming a co-propagating two-color bright-bright pulse pair in the waveguide, characteristic of the two-color soliton compression. In addition to pulse shortening, we also observe significant peak power enhancement.

To confirm the solitonic nature of the compression mechanism, we first turn to the soliton solutions of the coupled wave equations describing phase-mismatched second-harmonic generation (see Supplementary Section 2.1). When the group velocity dispersion (GVD) sign is the same for both the fundamental and second harmonic waves, there exists a well-known family of bright-bright soliton solutions\cite{he1996simultaneous}. Assuming 0 GVM, the shapes of the normalized fundamental and second-harmonic soliton envelopes, $a_\omega(\xi)$ and $a_{2\omega}(\xi)$ respectively, as a function of the dimensionless time coordinate \(\xi\) are approximately given by\cite{sukhorukov2000approximate}:

\begin{subequations}
\begin{equation}
a_{\omega}(\xi) = a_{\omega,0}\sech^p(\frac{\xi}{p}),\label{Eq3a}
\end{equation}
\begin{equation}
a_{2\omega}(\xi) = a_{2\omega,0}\sech^2(\frac{\xi}{p}),\label{Eq3b}
\end{equation}\label{Eq1}
\end{subequations}

\noindent where the parameters $p$, $a_{\omega,0}$, and $a_{2\omega,0}$ may be found using the equations:

\begin{subequations}
\begin{equation}
    p = \frac{1}{a_{2\omega,0} - 1},\label{Eq4a}
\end{equation}
\begin{equation}
    a_{\omega,0}^2 = \frac{\alpha a_{2\omega,0}^2}{a_{2\omega,0} - 1},\label{Eq4b}
\end{equation}
\begin{equation}
    \alpha = \frac{4(a_{2\omega,0}-1)^3}{2 - a_{2\omega,0}}.\label{Eq4c}
\end{equation}\label{Eq2}
\end{subequations}

Here, $a_{\omega,0}$ and $a_{2\omega,0}$ represent the normalized amplitudes of the fundamental and second-harmonic solitons. From this, we see that the soliton shape is completely determined by $\alpha = \abs{\frac{\beta_\omega^{(2)}}{\beta_{2\omega}^{(2)}}}(2+\frac{\Delta k}{\beta})$, where $\beta_\omega^{(2)}$ and $\beta_{2\omega}^{(2)}$ are respectively the GVD of the fundamental and second harmonic, $\Delta k$ is the phase mismatch, and $\beta$ represents shifts in the phase velocity due to the nonlinear interaction. This solution exhibits good agreement with the exact soliton solution, capturing both the behavior of the soliton amplitudes and tails. Furthermore, the solution shape asymptotes to that of the soliton in the cascading limit\cite{buryak1994spatial} where $\alpha \gg 1$ and precisely captures the known exact soliton solution\cite{karamzin1974nonlinear} with $\alpha = 1$.

\begin{figure}[hp!]
\centering\includegraphics[width=16cm]{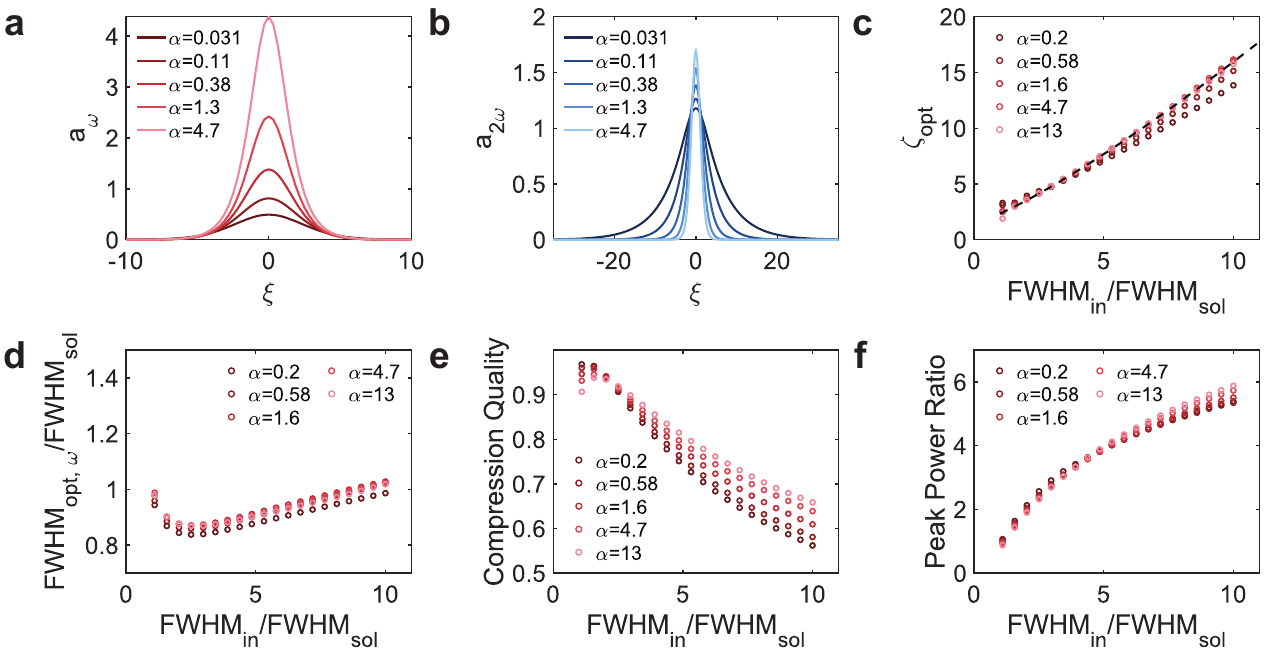}
\caption{{\bf Scaling behaviors of two-color soliton pulse compression.} {\bf a}, Soliton solutions of the fundamental wave for varying $\alpha$. {\bf b}, Corresponding soliton solutions for the second harmonic. {\bf c}, Optimum $\zeta$ for achieving compression. A fit is given by the dashed black line. {\bf d-f}, Scaling behaviors for varying $\alpha$ of the {\bf d}, fundamental FWHM,  {\bf e}, compression quality, and {\bf f}, fundamental peak power ratio at $\zeta_{opt}$. FWHM, full-width at half-maximum.}\label{fig2}
\end{figure}

In addition to this approximate analytic solution, we compute the soliton solution branch using numerical continuation (see Supplementary Section 2.2). Several examples of bright solitons for various $\alpha$ values are shown in Figs. \ref{fig2}a-b. As expected from equation \ref{Eq4b}, the amplitude of the normalized fundamental wave (Fig. \ref{fig2}a) is significantly larger than that of the second harmonic (Fig. \ref{fig2}b) for large $\alpha$ and vice versa for small $\alpha$.

Having to this point neglected walk-off, we next consider the effects of GVM, $\Delta\beta^{'}$, on the soliton solution. Here, we find that a soliton solution exists only in the stationary regime\cite{bache2007scaling} with $\abs{\beta_{2\omega}^{(2)}\beta}(2+\frac{\Delta k}{\beta})\geq\frac{(\Delta\beta^{'})^{2}}{2}$. This has presented a large challenge for achieving two-color soliton compression bulk systems with limited control over the dispersion, as overcoming the intrinsic GVM in the material has required operation in the cascading limit with large phase mismatch, limiting the power in the generated second-harmonic wave.

With the soliton solution in hand, we next use it to better understand the compression behavior. By investigating its stability, we find that the soliton is a saddle point with respect to the pulse amplitude, phase, and pulse width (see Supplementary Section 2.3). Thus, during the compression process, the fundamental and second-harmonic pulses approach the soliton solution, near which the evolution of the two waves is slow, and then are observed to again broaden. One must therefore optimize the length of the waveguide to achieve an optimally compressed pulse.

To determine this optimum length, we simulate the pulse evolution as a function of the normalized propagation coordinate, $\zeta = \abs{\beta}z$ for a variety of input parameters (see Methods). We then define the optimum distance, ${\zeta_{opt}}$, at which point the minimum pulse width is achieved for the fundamental wave. Figure \ref{fig2}c shows $\zeta_{opt}$ as a function of the ratio of the input pulse FWHM to that of the fundamental soliton, FWHM\textsubscript{in}/FWHM\textsubscript{sol}. As can be seen, the scaling behavior is similar for all values of $\alpha$ and nearly identical for $\alpha \geq 1$. By fitting the $\alpha \geq 1$ data (dashed, black line), we arrive at the following design heuristic:

\begin{equation}
\zeta_{opt} = 1.49 + 0.86\bigl(\frac{\text{FWHM\textsubscript{in}}}{\text{FWHM\textsubscript{sol}}}\bigr)^{1.23}.\label{Eq5}
\end{equation}

From these simulations, we may also study several key properties of the compressed pulse at the point $\zeta_{opt}$. To begin, we analyze the FWHM of the fundamental wave, FWHM\textsubscript{opt,\textomega}, and we compare it to FWHM\textsubscript{sol}. The results are shown in Fig. \ref{fig2}d. For all simulated input pulse widths, the width of the compressed pulse is within 20\% of the soliton width, with the pulses under-shooting the soliton pulse width for smaller values of FWHM\textsubscript{in}/FWHM\textsubscript{sol}.

A second parameter of interest is the compression quality\cite{bache2007scaling}, which is a measure of how well the energy remains localized in the pulse following compression. Here, it is defined as the ratio between the combined energy of the output fundamental and second-harmonic pulses and the input pulse energy. For the output, the energy is calculated from the pulse FWHM and amplitudes, assuming a sech-shaped pulse profile. As expected, the compression quality, shown in Fig. \ref{fig2}e, is higher for inputs with a FWHM closer to the soliton FWHM. However, a compression quality greater than 0.5 is observed even for the highest simulated ratio of FWHM\textsubscript{in}/FWHM\textsubscript{sol} = 10.

Finally, we are interested in the peak power enhancement provided by the compression mechanism, as an important benefit of the compressed pulses is their ability to drive nonlinear optical phenomena requiring large peak powers. The ratio between the peak power in the fundamental output and the input is plotted in Fig. \ref{fig2}f. Again, the trend is similar for all values of $\alpha$, with significant peak power enhancement observed for all simulated values of the input pulse FWHM.

These observations have several important consequences for the design of two-color soliton pulse compression systems. Firstly, since the compressed pulse exhibits a pulse width similar to that of the soliton solution and furthermore retains most of the input energy, the soliton solution given by equations \ref{Eq1} and \ref{Eq2} can be used to estimate the compressed pulse profile. Secondly, control over the dispersion parameters offers new opportunities for two-color pulse compression compared to previous demonstrations in the cascading limit with large \(\Delta k\), including operation in the small \(\alpha\) regime and compression with small \(\Delta k\). This can allow two-color compression with a variety of resultant pulse shapes and peak power ratios between the fundamental and second harmonics. Further discussion around the design of two-color soliton compression systems based on the presented theoretical framework may be found in Supplementary Section 2.5.

\subsection*{Device Design}

Based on these principles, we design a device for experimentally demonstrating two-color pulse compression to the near-single cycle regime. In our design, we aim to operate away from the cascading limit, in the small $\alpha$ regime. Furthermore, we seek to operate with low pump pulse energies on the order of a few pJ, as may be achieved by integrated ultrafast sources\cite{guo2023ultrafast, yu2022integrated}. The designed waveguide (see Supplementary Section 1.4) has a fundamental and second-harmonic GVD of $\beta_{\omega}^{(2)}$ = 9.2 fs\textsuperscript{2}/mm and $\beta_{2\omega}^{(2)}$ = 141 fs\textsuperscript{2}/mm, respectively, as well as a GVM of $\Delta\beta^{'}$ = 27 fs/mm between the two waves.

\begin{figure}[hp!]
\centering\includegraphics[width=16cm]{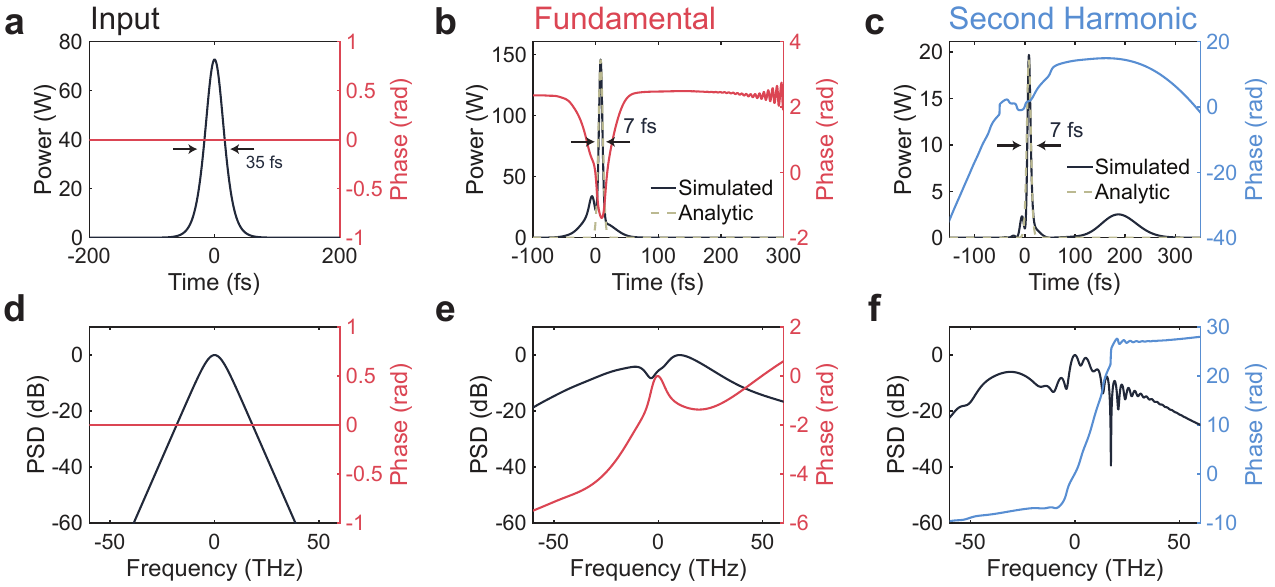}
\caption{{\bf Simulation of designed single-cycle pulse compressor.} {\bf a}, Input, {\bf b}, output fundamental, and {\bf c}, output second-harmonic pulses. {\bf d}, Corresponding input, {\bf e}, output fundamental, and {\bf f}, output second-harmonic spectra. Dashed, tan lines show the pulse profiles predicted from soliton theory.}\label{fig3}
\end{figure}

With these parameters, and considering our transform-limited input pulse width of 35 fs as well as a phase mismatch of $\Delta k$ = -4 rad/mm, we find that the soliton solution has a FWHM of 8 fs for an input energy of 3 pJ. This is nearly single-cycle for the fundamental wave at 143.5 THz. The corresponding normalized parameters in our system are $\beta = -1.02$ rad/mm and $\alpha = 0.39$. The optimum waveguide length, $L$, is then found using equation \ref{Eq5} to be 6.5 mm, the designed length for our nanophotonic device.

Simulation results for our designed device parameters are plotted in Fig. \ref{fig3}. The input is taken to be a 2.9-pJ, 35-fs sech-shaped pulse at 2090 nm, with temporal and spectral profiles shown in Figs. \ref{fig3}a and \ref{fig3}d. The fundamental output in time domain is shown in Fig \ref{fig3}b. The pulse profile is shown in dark gray, with the corresponding phase shown in red. Overlaid is the soliton solution given by equations \ref{Eq1} and \ref{Eq2} (tan, dashed line), exhibiting very good agreement. We normalize the peak power of the analytic solution to the peak power of the simulation, but we emphasize here that the soliton shape is otherwise unaltered. The pulse FWHM is 7 fs, close to the theoretical value (8 fs). This also equates exactly to a single cycle at the carrier frequency. Despite the inclusion of loss in the simulated waveguide, we additionally observe an approximately two-fold peak power enhancement at the fundamental. The corresponding carrier-free spectrum is shown in Fig. \ref{fig3}e and is characterized by a relatively flat phase across the entirety of the broadband spectrum.

The second-harmonic output is shown in Figs. \ref{fig3}c and \ref{fig3}f. As with the fundamental, the soliton solution is overlaid in the time domain plot (Fig. \ref{fig3}c). In this case, we perform no additional normalization, preserving the predicted peak power ratio of about 7.6:1 from the analytic solution. Yet, the agreement is again excellent. As with the fundamental, the FWHM is 7 fs. Although we operate in the soliton regime, the presence of some walk-off leads to the small secondary lobe at around 175 fs = $\Delta\beta^{'} L$. Correspondingly, we see that the carrier-free spectrum has more structure than for the fundamental wave, though the low-frequency side is observed to be smooth and to exhibit a fairly flat phase.

\subsection*{Experimental Results}

\begin{figure}[hp!]
\centering\includegraphics[width=16cm]{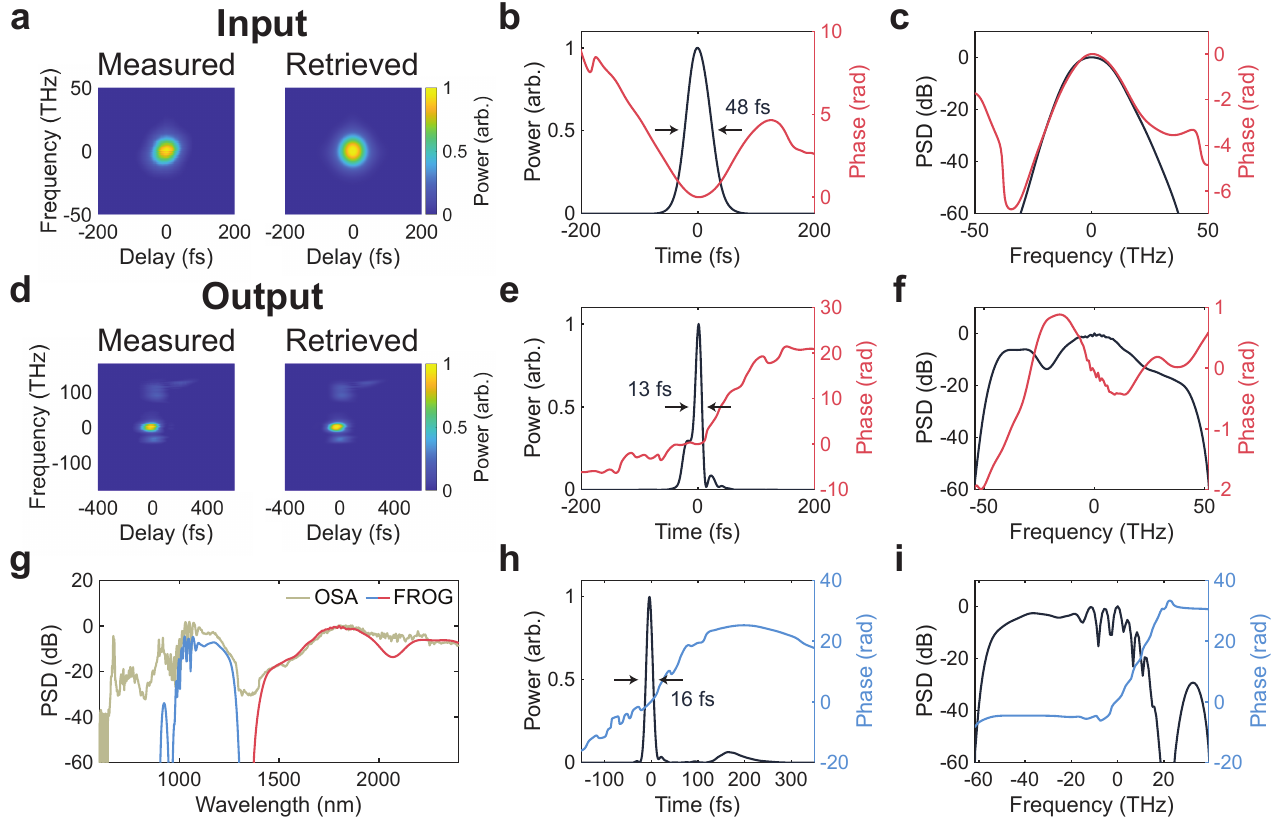}
\caption{{\bf Experimental quadratic soliton compression.} {\bf a}, Measured and retrieved SHG FROG traces of fundamental input pulse. {\bf b}, Input pulse temporal profile and {\bf c}, spectrum. {\bf d}, Measured and retrieved X-FROG traces of compressor output. {\bf e}, Output temporal profile and {\bf f}, spectrum for the fundamental. {\bf g}, Reconstructed FROG spectrum overlaid with measured OSA spectrum. {\bf h}, Output temporal profile and {\bf i}, spectrum for the second harmonic. FROG, frequency-resolved optical gating; OSA, optical spectrum analyzer. FROG errors of 0.0032 and 0.0046 were measured for the SHG FROG and X-FROG, respectively.}\label{fig4}
\end{figure}

To experimentally demonstrate the pulse compression, we fabricate the designed device and then temporally characterize the input and output pulses with a frequency-resolved optical gating (FROG) system (see Methods). Figure \ref{fig4}a shows the measured and retrieved SHG FROG spectrograms for the input pulse. Qualitatively, we see good agreement, corresponding to a reasonable FROG error of 0.0032. The reconstructed pulse and spectrum are shown in Figs. \ref{fig4}b-c. We observe a small amount of anomalous chirp due to propagation through various optical elements on the way to the chip setup, with the dominant contribution coming from a variable ND filter used for adjusting the input power.

The measured and retrieved X-FROG traces for the device output at a pump power of 3 pJ are shown in Fig. \ref{fig4}d. Again, good qualitative agreement is observed along with a reasonable FROG error of 0.0046. As further confirmation of the FROG performance, we compare the retrieved FROG spectrum with a secondary measurement on an optical spectrum analyzer (OSA). The result is shown in Fig. \ref{fig4}g, exhibiting good agreement across the entire spectrum. The largest discrepancy around 2090 nm is due to the presence of higher-order spatial modes, which are captured by the OSA but temporally gated by the X-FROG measurement due to their propagating at a different group velocity compared to the fundamental mode (see Supplementary Section 1.4). The slight under-estimation of power on the short-wavelength side of the spectrum and cut-off around 950 nm is predominantly due to a combination of the phase-matching bandwidth of the nonlinear crystal used in the FROG and the frequency response of a short-pass filter used at the FROG output to block residual light from the strong gate beam which can otherwise saturate the spectrum (see Supplementary Information 1.1). Finally, the discontinuity in the center of the FROG spectrum is due to the limited SNR of the FROG measurement.

The recovered fundamental pulse and spectrum are plotted in Figs. \ref{fig4}e and \ref{fig4}f, respectively. The pulse and spectrum exhibit qualitatively very similar behavior to the simulation, verifying the two-color soliton compression mechanism. The spectrum is broad and largely unstructured, besides a central dip, and the spectral phase exhibits only slow variation. The FWHM of the pulse is measured to be 13 fs, corresponding to less than two optical cycles for the fundamental carrier. Likewise, the recovered second-harmonic pulse and spectrum are plotted in Figs. \ref{fig4}h and \ref{fig4}i, respectively. Again, there is good agreement with the simulation. The pulse FWHM is also measured to be 16 fs. Furthermore, a small bump is observed in the vicinity of 175 fs as expected. Like the simulation, the spectrum is more structured than the fundamental but exhibits a flat phase and amplitude on the low-frequency side. While the measured pulses agree qualitatively well with the simulations, we finally note that discrepancies in the experimentally measured FWHM and simulation arise due to the impact of the chirp on the input pulse, higher order dispersion in the waveguide, and limitations in our current measurement setup (see Supplementary Sections 1.3 and 2.6).

\subsection*{Towards Single-Cycle Synthesis}

\begin{figure}[hp!]
\centering\includegraphics[width=16cm]{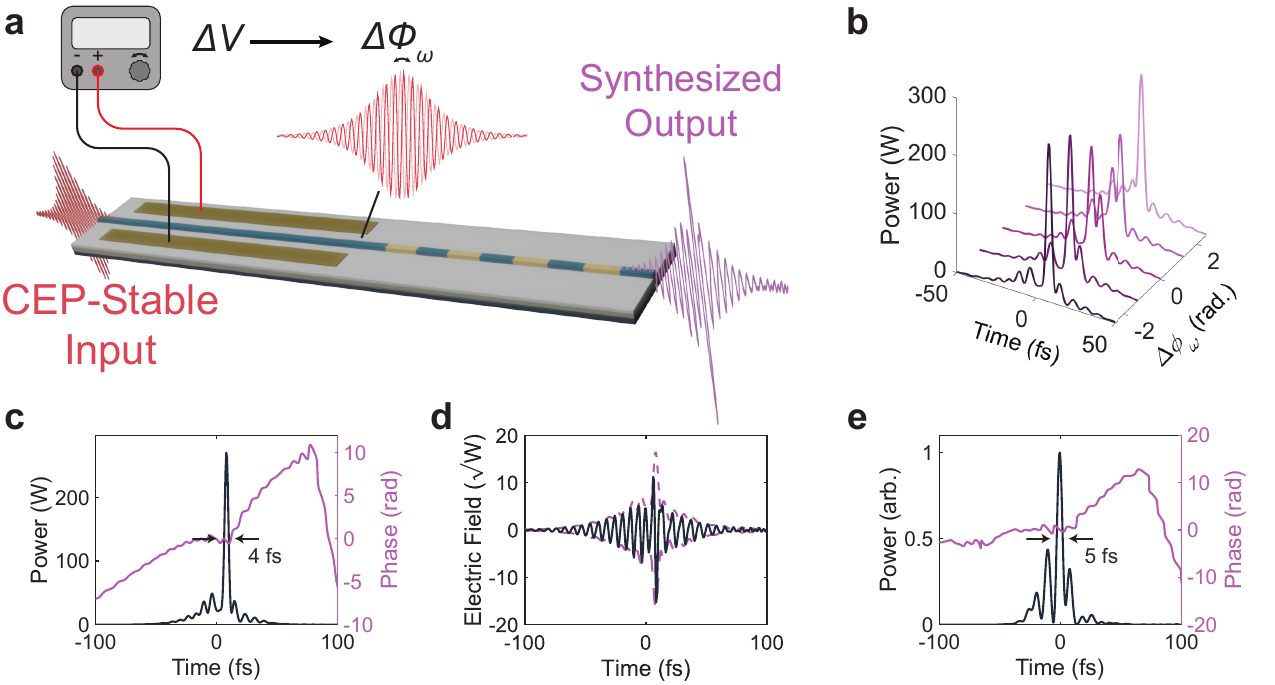}
\caption{{\bf Towards integrated single-cycle pulse synthesizers.} {\bf a}, Proposed nanophotonic circuit architecture for single-cycle pulse synthesis. {\bf b}, Simulated waveforms that may be achieved through manipulation of the input envelope phase, $\Delta\phi_{\omega}$. {\bf c}, Simulated synthesized single-cycle waveform and {\bf d}, corresponding electric field. {\bf e}, Expected waveform from synthesis of experimentally measured pulses.}\label{fig5}
\end{figure}

One unique feature of the two-color soliton compression is the opportunity it presents for facilitating on-chip single-cycle pulse synthesis. By manipulating the relative phase $\phi_{\omega} - \phi_{2{\omega}}$ of the two distinct harmonics, their combination can provide a variety of ultrashort waveforms. Interestingly, as the soliton solution occurs for a fixed phase relationship, $2\phi_{\omega}-\phi_{2\omega} = 0$ (see Supplementary Section 2.3), the relative phase between the two co-propagating harmonics may be manipulated through control of the envelope phase of the input. Thus, with a carrier-envelope phase (CEP)-stabilized input and relatively few additional components on-chip, an integrated single-cycle synthesizer may be envisaged (Fig. \ref{fig5}a). In our proposal, a voltage supplied to an integrated electro-optic modulator is used to directly tune the CEP of the input pulse\cite{gobert2011carrier}.

Figure \ref{fig5}b shows several examples of simulated synthesized pulse profiles as a function of the input pulse phase shift, $\Delta\phi_{\omega}$, for the parameters of Fig. \ref{fig3}. Figure \ref{fig5}c shows an example of a single-cycle pulse which may be realized through such a scheme, with a pulse FWHM of 4 fs and a combined carrier of 159 THz. The single-cycle nature of the pulse is highlighted through the plot of the corresponding electric field (solid line) and electric field envelope (purple, dashed line) shown in Fig. \ref{fig5}d. Additionally, we consider the synthesized pulse that may be realized using the experimentally measured 13 fs and 16 fs pulses, as plotted in Fig. \ref{fig5}e. Through overlapping and manipulating the relative phases of the measured traces, we observe that an ultrashort pulse with a FWHM of 5 fs may already be realized.

Finally, we note that typical intensities required for entering the regime of extreme nonlinear optics\cite{wegener2006extreme} are on the order of 10\textsuperscript{12} W/cm\textsuperscript{2}. With mode areas of $\sim 2$ \textmu m\textsuperscript{2} in the waveguide for the synthesized pulse, this requires peak powers on the order of 10 kW, whereas our current proposal using pJ pump pulses exhibits peak powers on the order of 100 W. However, with the continued development of nanophotonic sources, achieving the required 100-pJ pulse energies on chip may soon be feasible.

\section*{Discussion}

To summarize, we have demonstrated two-color soliton pulse compression in lithium niobate nanophotonics requiring only a few pJ of pump pulse energy. The experimentally measured fundamental pulse duration of 13 fs corresponds to less than two optical cycles of the carrier. Our results exhibit good agreement with theoretical models based on the analytic soliton solutions of the waveguide and numerical simulations of the coupled wave equations. We have further shown how the intrinsic phase relationship between the co-propagating fundamental and second-harmonic waves may be directly leveraged for single-cycle pulse synthesis. The compression mechanism may also be extended to longer pump pulses, making it compatible with integrated sources (see Supplementary Section 2.7). Taken together, our results offer a holistic design framework for achieving two-color soliton compression in quadratic media and illuminate the great potentials of this technique for realizing a new generation of scalable single-cycle pulse generators, enabling many applications in ultrafast integrated photonics. For example, the resultant ultrashort pulses can be leveraged to realize extremely high bandwidth information processing systems\cite{li2025all}. Furthermore, as higher-energy integrated pulsed sources become available, the compressed pulses may be used to drive high-harmonic generation as a compact source for high-resolution imaging and lithography\cite{goulielmakis2022high}.

\section*{Methods}

\subsection*{Numerical Simulation}

Simulations of the designed waveguide presented in the main text are performed using the Fourier split-step method to solve the coupled wave equations:\cite{jankowski2021dispersion,boyd2008nonlinear}, written as:

\begin{subequations}
\begin{equation}
\frac{\partial A_\omega}{\partial z} = -i\kappa A_{2\omega}A_\omega^*e^{-i\Delta kz}-\frac{i\beta_\omega^{(2)}}{2}\frac{\partial^2 A_\omega}{\partial t^2} - \frac{\alpha_{\omega}}{2},\label{CWEqa}
\end{equation}
\begin{equation}
\frac{\partial A_{2\omega}}{\partial z} = -i\kappa A_\omega^2e^{i\Delta kz}-\Delta\beta^{'}\frac{\partial A_{2\omega}}{\partial t}-\frac{i\beta_{2 \omega}^{(2)}}{2}\frac{\partial^2 A_{2 \omega}}{\partial t^2} - \frac{\alpha_{2\omega}}{2},\label{CWEqb}
\end{equation}
\end{subequations}

\noindent where $A_\omega(z,t)$ and $A_{2 \omega}(z,t)$ represent the amplitudes of the fundamental and second-harmonic waves at frequencies $\omega$ and $2 \omega$, respectively, normalized such that the instantaneous power in each wave is given by $\abs{A_j}^2, j\epsilon\{\omega, 2\omega\}$. The time coordinate is defined such that the reference frame is co-moving at the group velocity of the fundamental wave. $\kappa = \frac{\sqrt{2\eta_0}\omega d_{eff}}{n_\omega \sqrt{A_{eff} n_{2\omega}c}}$ is the nonlinear coupling coefficient, where $d_{eff}$ is the effective nonlinearity, $n_j$ is the refractive index of wave $j$, $A_{eff}$ is the effective mode area as defined in ref. \cite{jankowski2021dispersion}, $c$ is the speed of light, and $\eta_0$ is the impedance of free space. The GVM is given by $\Delta \beta^{'} = \frac{1}{v_{g,2\omega}} - \frac{1}{v_{g,\omega}}$, where $v_{g,j}$ is the group velocity of wave $j$, and $\beta_{j}^{(2)}$ is the GVD of the $j^{th}$ wave. Finally, the loss coefficient of wave $j$ is given by $\alpha_{j}$. Additional simulations including higher order dispersion can be found in Supplementary Section 2.6.

For finding the scaling laws of the soliton solution, simulations are performed using the normalized coupled wave equations, omitting loss:

\begin{subequations}
\begin{equation}
-i\text{sgn}(\beta)\frac{\partial a_\omega}{\partial \zeta} = \frac{\partial^2 a_\omega}{\partial \xi^2} - a_\omega + a_{2\omega}a_\omega^{*},\label{normCWeqa}
\end{equation}
\begin{equation}
-i\text{sgn}(\beta)\sigma\frac{\partial a_{2\omega}}{\partial \zeta} = \frac{\partial^2 a_{2 \omega}}{\partial \xi^2} + i\delta \frac{\partial a_{2\omega}}{\partial \xi} - \alpha a_{2 \omega} + \frac{a_\omega^2}{2}.\label{normCWeqb}
\end{equation}
\end{subequations}

The normalization procedure may be found in Supplementary Section 2.1. In our analysis, we take sgn$(\beta)$ = 1 for simplicity, though the conclusions are equivalent for sgn$(\beta)$ = -1. In the main text simulations, we further set the normalized walk-off parameter, $\delta$, to 0, and we consider only $\sigma = \alpha/10$ as $\alpha$ is the dominant parameter in determining the compression behavior. Further analysis on the impact of these two additional parameters may be found in Supplementary Section 2.2.

\subsection*{Device Fabrication}

We fabricate the device following the procedure described in ref. \cite{ledezma2022intense}. It is fabricated on X-cut MgO-doped thin-film lithium niobate on a SiO$_2$/Si substrate (NANOLN). To achieve the periodic poling, we pattern Cr poling electrodes using lift-off. By applying a voltage across the electrodes, we periodically flip the ferroelectric domains. Following poling, we etch the waveguides using Ar-milling with hydrogen silsesquioxane (HSQ) as the etch mask. Finally, we mechanically polish the waveguide facets to enable end-fire coupling into the devices. Further information on the device design and characterization may be found in Supplementary Section 1.4.

\subsection*{Experimental Procedure}

The experimental pulse measurements are conducted using a home-built FROG\cite{trebino2012frequency} system which utilizes a 50-\textmu m BBO crystal in a non-collinear geometry for broadband type-I phase matching between the signal and gate pulses. The FROG reconstruction is done using the principle components generalized projections algorithm\cite{kane1998real,kane1999recent,delong1996practical,Byrnesfrogalgorithm,Wyattsfrogalgorithm}. As an input to the compression device, we utilize 35-fs transform-limited pulses from a bulk degenerate optical parametric oscillator. The pulses are characterized using a second-harmonic generation (SHG) auto-FROG geometry. By contrast, the low-power output pulses from the chip are measured using a cross-FROG (X-FROG) geometry, gated by pulses generated from a commercial mode-locked laser. The 103-fs gate pulses are also characterized using a SHG auto-FROG geometry. Additional information regarding the experimental setup and FROG processing may be found in Supplementary Sections 1.1-1.3.

\printbibliography

\section*{Acknowledgments}

The device nanofabrication was performed at the Kavli Nanoscience Institute (KNI) at Caltech. The authors gratefully acknowledge support from DARPA award D23AP00158, ARO grant no. W911NF-23-1-0048, NSF grant no. 2408297, 1918549, AFOSR award FA9550-23-1-0755, the Center for Sensing to Intelligence at Caltech, the Alfred P. Sloan Foundation, and NASA/JPL. N.E. acknowledges support from the Belgian American Educational Foundation (B.A.E.F.) and the European Union’s Horizon Europe research and innovation programme under the Marie Skłodowska-Curie Grant Agreement No. 101103780.

\paragraph{Author Contributions:} R.M.G. and A.M. conceived the idea and designed the experiments. R.M.G. performed the simulations. R.M.G. performed the theoretical analysis with input from N.E. and L.L. M.S. designed the device with input from R.S., J.W., and R.M.G. R.M.G. and M.S. performed the experiments. R.M.G., T.Z., and R.C. developed and tested the FROG measurement system. M.S., J.W., and S.Z. fabricated the device. R.S. and L.L. fabricated earlier device iterations. A.M. supervised the project.

\paragraph{Competing Interests:}
R.S., R.M.G., and A.M. are inventors on a U.S. provisional patent application filed by the California Institute of Technology (application number 63/532,648). R.S., L.L., and A.M. are involved in developing photonic integrated nonlinear circuits at PINC Technologies Inc. R.S., L.L., and A.M. have an equity interest in PINC Technologies Inc.

\paragraph{Data Availability:}
The data used for generation of the figures within this manuscript and other findings of this study are available upon request from the corresponding author.

\paragraph{Code Availability:}
The code used for simulation and plotting of results is available upon request from the corresponding author.

\end{document}